\documentclass[prd,showpacs,amsmath,amssymb,twocolumn,superscriptaddress]{revtex4}

\usepackage{bm,amsmath}
\usepackage[dvips]{graphicx}

\newcommand{\qq}{\mathbf{q}}
\newcommand{\pp}{\mathbf{p}}

\newcommand{\rr}{\mathbf{r}}

\newcommand{\be}{\begin{equation}}
\newcommand{\ee}{\end{equation}}
\newcommand{\bea}{\begin{eqnarray}}
\newcommand{\eea}{\end{eqnarray}}
\newcommand{\ba}{\begin{align}}
\newcommand{\ea}{\end{align}}

\newcommand{\rme}{{\rm e}}
\newcommand{\rmi}{{\rm i}}

\usepackage{amssymb}

\begin{document}

\title{Interaction of gravitational waves with matter}
\author{A.\ Cetoli}
\affiliation{Department of Physics, Ume\aa ~University, SE-90187 Ume\aa , Sweden}
\affiliation{New Zealand Institute for Advanced Study and Centre for
  Theoretical Chemistry and Physics, Massey University, Private Bag
  102904 NSMC, Auckland 0745, New Zealand}
\author{C.\ J.\ Pethick}
\affiliation{The Niels Bohr International Academy, The Niels Bohr Institute, Blegdamsvej 17, DK-2100 Copenhagen \O, Denmark}
\affiliation{NORDITA, Roslagstullsbacken 23, SE-10691 Stockholm, Sweden }
\begin{abstract}
We develop a unified formalism for describing the interaction of
gravitational waves with matter that clearly separates the effects of
general relativity from those due to interactions in the matter.
Using it, we derive a general expression for the dispersion of
gravitational waves in matter in terms of correlation functions for
the matter in flat spacetime. The self energy of a gravitational wave
is shown to have contributions analogous to the paramagnetic and
diamagnetic contributions to the self energy of an electromagnetic
wave.  We apply the formalism to some simple systems - free particles, an interacting scalar field,
and a fermionic superfluid.
\end{abstract}

\maketitle

\section{Introduction} \label{sec:intro}
The interaction of gravitational waves with matter is important in a
number of different contexts.  One is in connection with the
continuing quest to detect gravitational waves experimentally.
Pioneering experiments were carried out using large metal bodies as
detectors \cite{weber1960} and this line of investigation has been
further pursued at many centers. 
These detectors were designed to detect gravitational waves by
observing the resonant excitation of elastic modes of the bars and the
standard theory of such detectors uses elastic theory to calculate the
response \cite{landauVol7}.

 However, there are a number of
suggestions that the response of matter could be very different from
what is predicted on the basis of elasticity theory.  One is that the
direct coupling of the gravitational wave to electrons could enhance
the absorption cross section \cite{widom2003}.  For solids, the effect
of electronic degrees of freedom has been taken into account within
the Fr\"ohlich model for the electron--phonon interaction, which does
not take into account explicitly the long-range character of the
Coulomb interaction, and the authors conclude that the effect of
including electron degrees of freedom explicitly is very small
\cite{branchina2004}.  A more recent suggestion is that a
superconducting metal would be a reflector of gravitational waves
because ions and superconducting electrons respond in different ways
to a gravitational wave, thereby creating a large electrostatic energy
that renders the superconductor ``stiff'' to the propagation of the
wave \cite{chiao2009}.  These proposals underscore the need for a
theory of the interaction of gravitational waves with matter that
treats coupling of the gravitational wave to matter on a unified
footing, takes into account the microscopic degrees of freedom, and
also is able to include the effects of interactions.

A second important area is the interaction of gravitational waves
with astrophysical matter.  Since much of this matter is diffuse and
weakly interacting, the common approach to this problem is to
calculate trajectories of free particles in the presence of the curved
spacetime produced by the gravitational wave.  A review of early work
on the dispersion of gravitational waves may be found in
Ref.\ \cite{grishchuck}.  The effects of electromagnetic fields are
included in some cases: for example, Servin, Brodin and Marklund
\cite{servin2002, servin2003} showed that a magnetic field can rotate the polarization of a gravitational wave. Their work suggests that the role of
the electromagnetic field is crucial to understanding the response of
a charged system.

The purpose of the present work is to develop a general formalism for
describing the interaction of gravitational waves with matter.  In
particular, we wish to separate the effects of general relativity from
those of calculating correlations in the matter.  Our approach is
modeled on the semiclassical theory of interaction of electromagnetic
fields with matter, in that we shall treat the gravitational radiation
classically.  However, the matter will be treated quantum
mechanically.  The response of the system is calculated in a
systematic way from a path-integral approach.  We find that there are
contributions to the response of matter to a gravitational wave that
are analogous to the paramagnetic and diamagnetic responses of a
conductor to an electromagnetic field, and we give general expressions
for them.  Earlier work on interaction of gravitational waves with
condensed matter \cite{widom2003,branchina2004} has generally focused
on the paramagnetic term, while in astrophysical applications the
diamagnetic term often dominates.  The formalism described in this
article provides an economical way of deriving results for simple
situations that have been considered earlier, while at the same time
being of sufficient generality to be applicable to interacting
many-body systems.

The paper is organized as follows: in Sec~\ref{sec:coupling} we develop the formalism for calculating the 
dispersion relation for a gravitational wave propagating in matter.   Section \ref{sec:free} treats the case of free particles, both non-relativistic and relativistic.  In Sec.\ \ref{sec:superfluids} we analyze the coupling of a gravitational wave to two interacting systems - a scalar boson field with a $\phi^4$ interaction (a Bose-Einstein condensate)  and a superfluid with paired fermions described by the Bardeen--Cooper--Schrieffer (BCS) theory.  We describe possible directions for future research in Sec.\ \ref{sec:conclusion}.

\section{Basic formalism} \label{sec:coupling}

In a gravitational wave, the metric tensor $g_{\mu\nu}(\mathbf{x},t)$
deviates from the Minkowski metric $\eta_{\mu\nu}={\rm
  diag}(1,-1,-1,-1)$ and we write
\bea 
g_{\mu\nu}(\mathbf{x},t) = \eta_{\mu\nu}+ h_{\mu\nu}(\mathbf{x},t)
\,,
\eea
where $h_{\mu\nu}(\mathbf{x},t)$ is the disturbance in the metric tensor.
In keeping with the general approach we adopt, the metric tensor will be treated as a classical quantity.
There is much freedom in the way in which the disturbance of the metric tensor is
described and, for gravitational waves, a convenient choice is the
transverse traceless (TT) gauge, in which $h$ has only spatial components, $\partial_\nu h^\nu_{\mu}=\partial_i h^i_j=0$, and
$h_{\nu}^\nu=0$ \cite{foot1}.  For gravitational waves, the quantity  $h_{\mu\nu}$ plays a role similar to that of the potential $A_\mu$ in electromagnetic theory.
In treating the effects of matter we shall assume that, in the absence of  gravitational waves, space-time is flat. This is a good approximation for wave numbers small compared with the scale of the curvature tensor \cite{servin2003}.

To describe the interaction of gravitational waves with matter, we generalize to gravitational waves the semiclassical theory of electromagnetic response \cite{altland}.
  The gravitational field is treated classically, but matter, including electromagnetic radiation, is treated quantum mechanically \cite{foot2}.  
The purely gravitational contribution to the action is
\be
S_{\rm grav}=\frac1{2\kappa}\int \sqrt{(-g)}R \approx \frac{1}{8\,\kappa} \int \frac{\partial h_{ij}}{\partial x_\sigma}
                          \,\frac{\partial h^{ij}}{\partial x^\sigma},
\ee
where the second expression is the leading contribution for small $h$, $\kappa = 8\pi G/c^4$ ($G$ being the Newtonian gravitational constant), $g=\det g_{\mu\nu}$, $R$ is the Ricci scalar and the
integrals are taken over space and time, $d^4x$.  

The contribution of matter to the effective action of the gravitational field due to matter is obtained by integrating over all possible paths for the quantum mechanical motion and is given by
\be
S_{\rm eff}(h) =\ln {\cal Z}(h).
\ee
Here the partition function ${\cal Z}(h)$ is given by \cite{foot3} 
\be
{\cal Z}(h)=\int{\cal D}(\bar \psi, \psi, A)\rme^{-S_m(\bar \psi, \psi, A, h)},  
\ee
where $A$ is the electromagnetic potential and the fields $\psi$ and $\bar \psi$ describe the other degrees of freedom of the matter.  By taking the integration over the complex time coordinate to run from 0 to $\rmi \beta$, where 
$\beta=1/T$ is the inverse temperature, one obtains a compact result which includes the effects of both quantum-mechanical and statistical averaging in the standard manner \cite{altland}.
The quantity $S_m$ is the contribution to the action from matter, and it may be written in the form
\be \label{actionmatter}
S_m=\int \sqrt{-g} \, {\cal L} ,
\ee    
where $\cal L$ is the Lagrangian function.
The equation for the deviation of the metric tensor is found from the extremum of the total effective action for the gravitational field, $S_{\rm grav}+S_{\rm eff}$, and has the form 
\bea \label{bareprop}
\Box h_{ij}= -4 \kappa \frac{\delta\ln {\cal Z}}{\delta h^{ij}}
\,,
\eea
where $\Box=
c^{-2}\,\partial_t^2-\nabla^2$ is the d'Alembertian operator. 
This is equivalent to the standard result $\Box h_{ij}= -2 \kappa T_{ij}$, where $T_{ij}$ is the energy--momentum tensor, since
\be
T_{ij} =2 \frac{\delta\ln {\cal Z}}{\delta h^{ij}}.
\ee
In the TT gauge, $h$ has no time components and therefore the indices run over the three spatial coordinates.  For definiteness, we shall consider a plane gravitational wave propagating along the $z$ axis, in which case the indices $i$ and $j$ can be either $x$ or $y$.

 For $h \rightarrow 0$, ${\delta\ln {\cal Z}}/{\delta h^{ij}}$ is independent of time and is irrelevant so far as gravitational waves are concerned. Expanding ${\delta\ln {\cal Z}}/{\delta h_{ij}}$ to first order in $h$ one finds
 \be \label{dressedprop}
\Box h_{ij}- \Sigma_{ij}^{kl}h_{kl}=0,
\ee
where 
\be \label{sigma}
\Sigma_{ij}^{kl}=-4 \kappa \frac{\delta^2\ln {\cal Z}}{\delta h^{ij}\delta h_{kl}}.
\ee
 Thus the quantity $ \Sigma_{ij}^{kl}$ plays the role of a self energy for the gravitational wave. For simplicity, we shall consider a medium that is isotropic, spatially homogeneous, and invariant under time reversal. It is then convenient to work with the quantities $h_+=(h_{xx}-h_{yy})/2$ and $h_\times=h_{xy}=h_{yx}$ that correspond to normal modes of the system, and these satisfy the equations
  \be \label{dressedpropplus}
\Box h_\times - 2\Sigma_{xy}^{xy}h_\times=0  \;\;\; {\rm and} \;\;\; \Box h_+ - 2\Sigma_{xy}^{xy}h_+=0,
\ee   
the factor of two being due to the fact
that in Eq.\ (\ref{dressedprop}) $kl$ can be both $xy$ and $yx$.

In order to find an expression for $\Sigma$ for small perturbation in the metric, we
expand the Lagrangian density in powers of $h$,
\begin{eqnarray} \label{Lcovcontra}
\sqrt{-g}\,{\cal L}
&=& {\cal L}^0
 + \left.\frac{\delta \sqrt{-g}\, {\cal L}}{\delta h^{ij}}\right|_{h=0}\,h^{ij}
\nonumber\\
&+& \frac12
    \left.\frac{\delta^2 \sqrt{-g}\, {\cal L}}{\delta h^{ij}\delta h_{kl}}\right|_{h=0}
          \,h^{ij}\,h_{kl}
 + O(h^3)
\,.
\end{eqnarray}
Quite generally, the energy--momentum or stress tensor is given by \cite{MisnerThorneWheeler}
\be \label{stress_deriv1}
T_{ij}= \frac{2}{\sqrt{-g}}\frac{\delta \sqrt{-g}\, {\cal L}}{\delta h^{ij}},
\ee
and therefore 
\be \label{stress_deriv}
\left.\frac{\delta \sqrt{-g}\, {\cal L}}{\delta h^{ij}}\right|_{h=0}\,h^{ij}
 = \frac{1}{2}\,T_{ij}^0\,h^{ij}
\,,
\ee
where $T_{ij}^0$ is the stress tensor in flat spacetime.

Since the self energy is a second functional derivative with respect to $h$, it is important that all quantities of second order in $h$ are calculated consistently.  In particular, care must be taken to distinguish upper and
lower indices, as one can see by noting that the condition
$g^{ik}\,g_{kj}=\eta_j^i$ implies that 
\begin{eqnarray}
h^{ij}\approx h_{ij} - h_{ik}\,h_{kj}
\,
\end{eqnarray}
to second order in $h$.  Moreover, remembering that $g_{ij}=
\eta^{ij}+h^{ik} + O(h^2)$, one finds
\bea 
T_{ij}&=& g_{ik}\,T^{kl}\,g_{lj}
\nonumber\\
          &\approx& T^{ij} + h^{ik}\,T^{kl}+ T^{ik}\,h^{kl}
\,.
\eea
After some calculation, the equation for the field is obtained:
\bea \label{prop} 
\Box h_{ij}
          &=&  
            2 \kappa\,\langle \delta T_{ij}^\mathrm{dia} \rangle
             +
            2 \kappa\,\langle \delta T_{ij}^\mathrm{para}  \rangle
\,,
\eea
where 
\bea
\langle \delta T_{ij}^\mathrm{dia} \rangle 
   = 
                \, \langle T^0_{ii}+ T^0_{jj}\rangle \,h_{ij}
              + 2\, \langle\frac{\delta^2 \sqrt{-g}\,{\cal L}}{\delta h_{ij}\delta h_{kl}}\rangle
                     \,h_{kl}
\,,
\label{dia}
\eea
is what we shall refer to as the ``diamagnetic'' contribution and, with arguments written out explicitly,
\begin{equation}
\langle \delta T_{ij}^\mathrm{para}(1) \rangle=\frac{\rm i}2\int d{\bf r}_2dt_2\,\theta(t_1-t_2)\langle
[T^0_{ij}(1),T^0_{kl}(2)] \rangle\,h_{kl}(2)
\label{para}
\end{equation}
is the ``paramagnetic'' contribution. 
 Here $\langle O \rangle=\int
D[\bar\psi,\psi,A]\, O\, \rme^{-S_m(\bar \psi, \psi, A, h=0)}$ denotes
the thermal average of the operator $O$ in flat spacetime, $[A,B]$ denotes the commutator, and $\theta(t)$ is the unit step function.   If $h_{kl}$ varies in time as $\rme^{-\rmi \omega t}$, one therefore finds
\be
\langle \delta T_{ij}^\mathrm{para}({\bf r}_1,t) \rangle = \frac12\int d\rr_2 \chi_{ij,kl}(\rr_1,\rr_2)\delta h_{kl}(\rr_2),
\ee
where 
\bea
\chi_{ij,kl}(\rr_1, \rr_2)
&=&
  \sum_n P_n\Bigg\{\frac{(T^0_{ij}({\bf r}_1))_{nm}(T^0_{kl}({\bf r}_2))_{mn} }{E_n-E_m +\omega +\rmi\eta} 
 \nonumber \\
&+&  \frac{(T^0_{kl}({\bf r}_2))_{nm}(T^0_{ij}({\bf r}_1))_{mn}}{E_n-E_m-\omega-\rmi \eta}   \Bigg\} ,
\label{paraFT}
\eea
where $P_n$ is the statistical weight of energy eigenstate $n$ for the system in flat space.

If the Lagrangian density is local in time, the diamagnetic term is
independent of frequency and therefore behaves as a mass term.  For matter described using a nonrelativistic framework, the Lagrangian density may be nonlocal in space, in which case the mass will depend on the wavevector.
The paramagnetic term depends on both frequency and wave vector and contains information about excited states of the matter.

We now examine the various contributions to the response of the stress tensor.
The first term on the right-hand side of Eq.\ (\ref{dia}) is proportional to the pressure of the matter and it is therefore independent of the frequency of the gravitational wave. If the Lagrangian density is local in time (as is usually the case),  the second term is also frequency-independent.  Thus both these terms behave as a mass term for the gravitational wave, which is why we refer to their total as the diamagnetic term.

Notice that the expectation value of the second derivative of the
Lagrangian in (\ref{prop}) vanishes identically for free particles,
leaving only terms proportional to the pressure; this is in accord
with the previous results of the relativistic literature
\cite{grishchuck,servin2003}. However, this term gives a contribution
to the dispersion relation in the presence of an interparticle
potential. For example, this contribution must be taken in account for calculating the self energy of a gravitational wave interacting with a scalar field (a Bose-Einstein condensate), as we
shown in Sec. \ref{sec:bec}.

The only contribution to the self energy that has an energy denominator and can thereby give rise to absorption of gravitational waves by matter is
the paramagnetic term.The expression
(\ref{prop}) is analogous to that for an
electromagnetic wave in a conductor, where the response of the matter
is evaluated self-consistently using linear response theory. 
The paramagnetic term does not contain
the current--current response, but an equivalent expression with the
stress tensor. For an electromagnetic wave the diamagnetic term is proportional
to the particle density, while for a gravitational wave  it contains terms proportional to the
pressure and terms proportional to the second derivative of the
Lagrangian density.  
In the equation for the field $h_{ij}$ the right hand side contains
only quantum mechanical and thermal averages of quantities in flat
space.  Thus in the present  approach, the effects of general relativity
  have been decoupled from the problem of solving the many-body
  problem for the matter. This is possible because the gravitational
fields are weak.
~\\

A number of properties of response functions at long-wavelength may be obtained by a consideration of conservation laws.  This has previously been done for the density, spin density and current responses in the context of Fermi liquid theory \cite{LeggettI, 
  pinesnozieres, baympethick, olsson1,olsson2} and we here apply these ideas to the stress-tensor response.

We consider the response of an initially uniform
medium to the application of a perturbation having the form
\bea
H_1=\int d{\bf x}{\cal O}_{\bf q} U_{-\qq}{\rm e}^{\rmi \qq\cdot \rr-\rm i\omega t}+h. c.,
\eea
where $U_{-\qq}$ is the strength of an applied external field and ``h.\ c.'' denotes the Hermitian conjugate.  If the operator $\cal O$ satisfies a local conservation law, in coordinate space the operator equation for the conservation law is
\be
\frac{\partial {\cal O}}{\partial t}+{\bm\nabla}\cdot {\bf j}^{\cal O}=0,
\ee
where ${\bf j}_{\cal O}$ is the operator for the corresponding current.  On taking matrix elements of the Fourier transform of this relation between energy eigenstates of the unperturbed system, which are labelled by $m$ and $n$, one finds
\be
\omega_{nm}({\cal O}_{\bf q})_{nm}= \qq \cdot ({\bf j}^{\cal O}_ \qq)_{nm}\, .
\ee
This equation demonstrates that, if $\cal O$ satisfies a local conservation law, and provided the corresponding current is not divergent in the long-wavelength limit, then matrix elements of $({\cal O}_{\bf q})_{nm}$ vanish for $\qq \to 0$ for all states for which $\omega_{nm}\neq 0$.  Expressed in other terms, this states that the only nonvanishing matrix elements of $({\cal O}_{\bf q})_{nm}$ are between states whose energy difference falls off at least as rapidly as $q$.  It is this observation when applied to the particle density, and in the case of translationally invariant systems also the particle current density, that lies behind the success of Landau Fermi liquid theory in providing a powerful way of parametrizing the properties of long-wavelength properties of normal Fermi liquids.  The calculations of response functions for interacting systems in Sec.\ \ref{sec:superfluids} will illustrate these general properties, but first we describe results for free particles.  In this case there are matrix elements of the stress tensor operator to states having nonzero excitation energy vanish in the long-wavelength limit because the momentum of a particle and its velocity are both conserved quantities and, consequently, the contribution of a particle to the stress tensor is conserved.

\section{Free particles} \label{sec:free}

In this section, we study the  response of a system of noninteracting particles to a gravitational wave. First, we consider free particles and derive simply results previously obtained by other methods in the astrophysical  literature. In addition, we explore two other systems where the quantum mechanical nature of the system is
relevant: the Bogoliubov theory of a Bose--Einstein condensate and the
BCS theory of superfluid fermions.

We begin by considering the case of a noninteracting, nonrelativistic
particles.  The diamagnetic contribution to the self energy of the
gravitational wave, Eq.\ (\ref{dia}) may be calculated simply for free
particles obeying either the Schr\"odinger equation or the
Klein--Gordon equation, since $\langle\cal L\rangle$ is zero and
therefore $\langle\delta^2 \sqrt{-g}\,{\cal L}/\delta h^2 \rangle= 0$
in the TT gauge, and the diamagnetic contribution contains only the
pressure $P=\langle T_{ii}\rangle$; as we shall show, for
nonrelativistic particles the diamagnetic contribution is smaller by a
factor $\langle v^2\rangle /c^2$, where $\langle v^2\rangle$ is the
mean square particle velocity.  Thus one finds
\bea \label{diamagnetic}
\omega^2
   &\approx& 
      c^2\,q^2 + \frac{32\pi GP}{c^2}  
\,, 
 \eea
to first order $G$ and first order in ${\langle v^2 \rangle}/{c^2}$.
For fermions at zero temperature the pressure is $P=
n\,p_F^2/5\,m$, where $p_F$ is the Fermi momentum; the dispersion
relation is therefore
\bea \label{fermions}
\omega^2
   &\approx& 
     c^2\,q^2 + \frac{32\pi}{5}Gmn\frac{ v_F^2 }{c^2}
\,.
 \eea
For an ideal gas obeying Maxwell--Boltzmann statistics, the pressure is $P= n\,T$, and the dispersion
relation becomes
\bea
\omega^2
   &\approx& 
    c^2\,q^2 + 32\pi\,Gmn\frac{ T }{m\,c^2}
      \label{classicaldispersion}
\,.  \eea 
We see that the dispersion relation depends on the ``Jeans'' frequency
$\omega_G=\,(G \,n\,m)^{1/2}$ characteristic of gravitational
collapse, reduced by a factor $\langle v^2 \rangle ^{1/2}/c$. The
result (\ref{classicaldispersion}) coincides with the result in the
literature \cite{servin2003, asseo1975, gayer1979}.

We turn now to the paramagnetic contribution.  In order to simplify
the discussion, it is convenient to consider the specific case of a
gravitational wave corresponding to a disturbance of
$h_{xy}=h_{yx}=h_\times$ propagating in the $z$-direction. The stress
tensor for free particles in flat space is given by
\be
  \hat{T^0}_{xy}(1) = \frac{1}{4\,m}\,(\nabla_1 - \nabla_{1'})_x
                                        \,(\nabla_1 - \nabla_{1'})_y
                                        \left.
                                        \,\hat{\psi}^\dagger(1')
                                        \,\hat{\psi}(1)
                                        \right|_{1'\rightarrow 1}
\,,
\ee
where $\hat{\psi}^\dagger(1)$ is the particle creation operator  and $\hat{\psi}(1)$ the annihilation operator at the
point $({\bf r}_1,t_1)$.  In a uniform medium, it is convenient to work with
the spatial Fourier transform of this quantity, which is given by 
\be
(\hat{T^0}_\qq)_{xy} = \sum_\pp \frac{p_x p_y}m \,{\hat
  a}^\dagger_{\pp-\qq/2} \hat a_{\pp +\qq/2} \,. 
  \label{stress_uniform}
\ee 

In the literature the response to a gravitational wave is often
studied by using the Vlasov equation to model the behavior of the
excitations in the system. In this section we first present the Vlasov
equation approach and then show that the same results may be obtained
by a quantum mechanical treatment.

In a homogeneous system, the perturbation in the stress tensor of a
non-interacting gas is given by
\bea
\delta T_{xy}^{\mathrm{para},0}({\bf q})
              &=& \int \frac{d\mathbf{p}}{(2\,\pi)^3} t^0_{xy} (\mathbf{p})
                                   \,
                                   \,\delta n_{\bf p}({\bf q})
\,,
\eea

where
\be
t^0_{xy} (\mathbf{p})= \frac{p_x\,p_y}{m}\,  \label{simple_stress}
\ee
is the stress tensor associated with a single particle,
and the Vlasov equation reads
\bea
\frac{\partial}{\partial t} \delta n_{\bf p}
+ 
\mathbf{v} \cdot \bm\nabla_{\bf r} \delta n_{\bf p}
-
(\bm\nabla_{\bf r} t^0_{xy}\,h_{xy}) \cdot \bm\nabla_{\bf p} n^0_{\bf p} = 0
\,.
\eea
Thus one finds
\be \label{nonint_stress}
\delta T^\mathrm{para}_{xy}
             = \chi_{xy,xy} \, h_{xy} ,
\ee             
 where           
 \be
    \chi_{xy,xy} =             - \int \frac{d\mathbf{p}}{(2\,\pi)^3} 
                   \frac{ \,p_x^2\,p_y^2}{m^2}
                    \,
                    \,\frac{\mathbf{q} \cdot \bm\nabla_{\bf p}n^0_{\bf p}}
                           {\omega - \mathbf{q} \cdot {\bf p}/m}
                 \, h_{xy}
\,,
\label{deltaTfree}
\ee 
is the transverse 
stress-tensor--stress-tensor  response function.

For gravitational waves, the frequencies of interest are approximately $cq$ and therefore for nonrelativistic particles one may expand the denominator in Eq.\ (\ref{deltaTfree}) and to leading order in $1/\omega^2 $ the result is
\bea
\delta T^\mathrm{para}_{xy}
             \simeq 
                 -\frac1{\omega^2} \int \frac{d\mathbf{p}}{(2\,\pi)^3} 
                   \frac{ \,p_x^2\,p_y^2}{m^2}
                    \,
                    \,\frac{(\mathbf{q} \cdot {\bf p})^2}{m^2}
                     \frac{\partial n^0_{\bf p}}{\partial \epsilon_p }
                                            \, h_{xy}
\,.
\eea
For nondegenerate particles, the distribution function is  Maxwellian and one finds
\bea
\chi_{xy,xy}
             &\simeq&  n\,\frac{T^2}m\frac{q^2}{\omega^2}
\,,
\eea
while for a Fermi gas at zero temperature
\bea
 \label{fermion_Sigma}
\chi_{xy,xy}
             &\simeq&\,
                  \frac{1}{35}\, nm v_F^4 \frac{q^2}{\omega^2}\, 
\,,
\eea
where in Eq. (\ref{fermion_Sigma}) we used the fact that $n=N/V=\nu
p_F^3/6\,\pi^2$, where $\nu$ is the number of degenerate internal
states of the particle, due to spin, isospin or other
symmetries. By including the paramagnetic term, the dispersion relation 
becomes
\bea \label{full_dispersion}
\omega^2
   &\approx& 
      c^2\,q^2 + \frac{32 \pi GP}{c^2} + 16\pi G\frac{\chi_{xy,xy}}{c^2}
\,.
 \eea

~\\

The results may also be obtained from a quantum-mechanical calculation based on Eq. (\ref{nonint_stress}). In the notation of second quantization, the stress tensor operator for a free particle system is
given by
\be
  \hat{T}_{xy}(1) = \frac{1}{4\,m}\,(\nabla_1 - \nabla_{1'})_x
                                        \,(\nabla_1 - \nabla_{1'})_y
                                        \left.
                                        \,\hat{\psi}^\dagger(1')
                                        \,\hat{\psi}(1)
                                        \right|_{1'\rightarrow 1},
\,.
\ee
where to simplify the notation we do not  write sums over internal states explicitly.
The expectation value of the stress tensor therefore reads
\be
 \langle \hat{T}_{xy}(1) \rangle
                           = -\frac{1}{4\,m}\,(\nabla_1 - \nabla_{1'})_x
                                        \,(\nabla_1 - \nabla_{1'})_y
                                        \left.
                                         \mathcal{G}(1,1')
                                        \right|_{1'\rightarrow 1^+}
\,,
\ee
where $ \mathcal{G}(1,1')= - \langle \hat{\psi}^\dagger(1')\,\hat{\psi}(1) \rangle$  is the (finite temperature) single-particle Green function. 
The paramagnetic response is given by 
\begin{widetext}
\bea
\langle \delta \hat{T}^\mathrm{para}_{xy}(1)\rangle 
&=& \rmi \int d\mathbf{r}_2 dt_2 \,\frac{1}{16\,m^2}\,(\nabla_1 - \nabla_{1'})_x
                                        \,(\nabla_1 - \nabla_{1'})_y
                                        \,(\nabla_2 - \nabla_{2'})_x
                                        \,(\nabla_2 - \nabla_{2'})_y
\nonumber \\
&~&~~~~~~~~~~~ \times               \,\langle \left[ \hat{\psi}^\dagger(1')
                                            \,\hat{\psi}(1),
                                            \hat{\psi}^\dagger(2')
                                            \,\hat{\psi}(2)\right] \rangle
                                                                    _{\begin{array}{c} 1'\rightarrow 1^+\\
                                                                               2'\rightarrow 2^+\end{array}}
                               \, h_{xy}(2)
\,,
\eea
which in Fourier space becomes
\bea   \label{gen_resp}
\langle \delta \hat{T}^\mathrm{para}_{xy}(\mathbf{q},\rmi \, \omega_n)\rangle
&=& -\,
        \sum_{\mathbf{p}, \omega_l}
                               \,\frac{p_x^2 p_y^2}{m^2}
                               \, \mathcal{G}(\mathbf{p}+\mathbf{q},\rmi \omega_l+\rmi \omega_n)\,\mathcal{G}(\mathbf{p},\rmi\omega_l)
                               \, h_{xy}(\mathbf{q},\rmi \omega_n)
\nonumber \\
&=& 
    -\,
    \sum_{\mathbf{p}}
                               \,\frac{p_x^2 p_y^2}{m^2}
                               \frac{n^0(p+q)-n^0(p)}
                                    {\rmi\,\omega_n - E_{p+q} + E_p}
                               \,\, h_{xy}(\mathbf{q},\rmi \omega_n)
\,, 
\eea
\end{widetext}
where the Matsubara frequencies are $\omega_n=2\pi nT$ and
$\omega_l=2l\pi T$ for bosons ($(2l+1)\pi T$ for fermions).  For $q
\ll \langle p\rangle$, Eq.\ (\ref{gen_resp}) reduces to
Eq.\ (\ref{deltaTfree}) obtained from the Vlasov equation.
~\\

We now comment briefly on the case of relativistic particles, and for
definiteness we shall consider particles described by the
Klein--Gordon equation.  As remarked above, for such particles the
diamagnetic response is given in terms of the pressure, just as for
nonrelativistic particles.  The calculations for the paramagnetic
response may be performed essentially as before and for $q\ll \langle
p\rangle$ the effect is to replace the mass $m$ by the ``relativistic
mass'' $m(1+(p/mc)^2)^{1/2}$.  In general, one cannot assume that the
particle velocity is small compared with $c$, and consequently the
response function has to be evaluated numerically.  Simple results may
be obtained for ultrarelativistic particles, since the particle
velocity is $c$ for all momenta.  In that case, the integrals over the
polar angle and the magnitude of the momentum decouple and one finds
\bea
\chi_{xy,xy} 
             &=&
                  \int \frac{d\mathbf{p}}{(2\,\pi)^3} 
                   \frac{ \,p_x^2\,p_y^2}{p^2/c^2}
                    \,
                    \,\frac{\mathbf{q} \cdot \bm\nabla_{\bf p}n^0_{\bf p}}
                           {\omega - \mathbf{q} \cdot {\hat{ \bf p}}\,c}  \nonumber\\
             &=& c\int_{-1}^{1}\frac{d\mu}2\frac{ \mu(1-\mu^2)^2}{s-\mu}\int_0^\infty\frac{4\pi p^4 dp}{(2\pi)^3}\frac{\partial n^0_{\bf p}}{\partial p}\nonumber \\
              &=&\frac23 P  \big( -\frac{16}{15} + \frac{10}{3} s^2 
\nonumber\\
&~&~~~- 2 s^4 +  s (s^2-1)^2 \ln\left[\frac{s+1}{s-1}\right]\big)
,
\eea 
where ${\hat \pp}=\pp/p$, $\mu={\hat \pp}\cdot\hat{\bf z}$ and
$s=\omega/cq$. The response function for the transverse components of
the stress tensor is finite for $s=1$, unlike the density and current
response functions, which have a logarithmic divergence.  The physical
reason for this is that particles moving in the direction of
propagation of the wave give vanishing contributions to the transverse
components of the stress tensor.  For $s=1$,
$\chi_{xy,xy}=8P/45$, which is of the same order as the
diamagnetic response. This result can be added to the diamagnetic contribution, Eq. (\ref{diamagnetic}),
to give the following dispersion relation for the ultrarelativistic case:
\bea
\omega^2 = c^2\,q^2 + \frac{1568 \pi G}{45 c^2}\,P
\,.
\eea
                 
  \section{Interacting systems} \label{sec:superfluids}
  
 In this section we consider two examples of interacting systems at zero temperature.  The first is an interacting boson field. Due to the interaction, this has a nonvanishing  diamagnetic contribution to the response to a gravitational wave.  Both it and the BCS superfluid have paramagnetic contributions due to excitation of two excitations with nonzero energy even at long wavelengths, and serve as an illustration of the general results described at the end of Sec.\ \ref{sec:coupling}.  
                 
\subsection{Interacting boson field} \label{sec:bec}

The Lagrange function for a
nonrelativistic boson field (a Bose--Einstein condensate) with a short-range interaction is
\bea
{\cal L}
&=&\frac{1}{2m}g^{ij}\nabla_i \hat{\psi}^\dagger \nabla_j \hat{\psi} +mc^2{\hat{\psi}}^\dagger\hat{\psi}+\frac{U_0}{2}(\hat{\psi}^\dagger)^2\hat{\psi}^2
\nonumber\\
&-&\frac{\rmi \hbar}{2}\left(  \hat{\psi}^\dagger \frac{\partial \hat{\psi}}{\partial t}
+ \frac{\partial \hat{\psi}^\dagger}{\partial t}\hat{\psi} \right),
\eea
where, in the Gross--Pitaevskii approach,$U_0=4\pi\hbar^2a/m$, $a$ being the scattering length for two-body scattering, is the strength of the effective two-body interaction. Therefore the spatial components of the stress tensor (\ref{stress_deriv1}) are
\bea\label{stress_bec}
\hat{T}_{ij}(\mathbf{x})
&=& -\frac{1}{4\,m}\,(\nabla_1 - \nabla_{1'})_i
                 \,(\nabla_1 - \nabla_{1'})_j
                 \left.
                 \,\hat{\psi}^\dagger(1')
                 \,\hat{\psi}(1)
                 \right|_{1'\rightarrow 1}
\nonumber\\
&+& \delta_{ij}\,\frac{U_0}2 \, \hat{\psi}^\dagger(\mathbf{x})^2 \hat{\psi}(\mathbf{x})^2
\,.
\eea

The last term in (\ref{stress_bec}) is the pressure due to the
interparticle interaction and in the Bogoliubov approximation, in
which $\hat\psi$ is replaced by a c-number $\sqrt{n_0}$ with $n_0$ the
condensate density, it becomes $n_0^2U_0/2$.  If depletion of the
condensate may be neglected, $n_0$ may be replaced by $n$ and the
result agrees with the one obtained from the thermodynamic relation
$P=n^2\partial (E/n)/\partial n$, where $E$ is the energy density.

The interaction does not contribute to the paramagnetic term in the dispersion relation, because the gravitational wave is transverse, but there is a diamagnetic term since
\bea
\langle\left.\frac{\delta^2 \sqrt{-g}\,{\cal L}}{\delta h_{ij}\delta h_{kl}}\right|_{h_{ij}=0}\rangle 
&=& \langle {\cal L}^0 \rangle \nonumber\\
&=& P\,,
\eea
where $P= U_0\,n^2/2$ is the pressure. The contribution $\delta
T^\mathrm{dia}$ gives then the dispersion relation
\bea
\omega^2
   &\approx& 
      c^2\,q^2 + 16 \, \pi\,G \,n\,m \,\frac{nU_0}{m\,c^2}
\,.
\eea

We now consider the paramagnetic term, which is not generally zero.
 In Fourier space the contribution to the stress tensor from the kinetic energy  may be written as
\bea
\hat{T}_{ij}(\mathbf{q},\omega)
&=& \frac{1}{m}\,\sum_{\mathbf p} (p+\frac{q}{2})_i\,(p+\frac{q}{2})_j
         \,\hat{a}_\mathbf{p}^\dagger\,\hat{a}_{\mathbf{p}+\mathbf{q}}
\nonumber\\
&=& \frac{1}{4\,m}\,\sqrt{n_0}\,q_i\,q_j\,(\hat{a}_\mathbf{q}
                                           +\hat{a}_{-\mathbf{q}}^\dagger)
\nonumber\\
&+&\frac{1}{m}\,\sideset{}{^{'}}{\sum}_\mathbf{p} (p 
                                             +\frac{q}{2})_i
                                             \,(p+\frac{q}{2})_j
         \,\hat{a}_\mathbf{p}^\dagger\,\hat{a}_{\mathbf{p}+\mathbf{q}}
\nonumber\\
&=& \frac{1}{m}\,\sideset{}{^{'}}{\sum}_\mathbf{p} p_i\,p_j
         \,\hat{a}_\mathbf{p}^\dagger\,\hat{a}_{\mathbf{p}+\mathbf{q}}
\,,
\eea
where the prime in the sum over $\bf p$ denotes that only the atoms
outside the condensate are taken in consideration. The condensate
contribution vanishes identically, because the only non-zero component
of $\bf q$ is in the $z$ direction. This component never appears in the
transverse response, i.e. $q_i=q_j=0$ for $i,j=1,2$.

Elementary excitations of the condensate are created by operators $\hat{\alpha}^\dagger_{\mathbf k}$ and destroyed by $\hat{\alpha}_{\mathbf k}$, which are related to the the particle creation and annihilation operators by 
the Bogoliubov transformation
\bea
\hat{a}_{\bf k} = u_{\bf k}\,\hat{\alpha}_{\bf k} - v_{\bf k}\,\hat{\alpha}^\dagger_{-{\bf k}}\, ,
  \nonumber\\
\hat{a}^\dagger_{\bf k} = u_{\bf k}\,\hat{\alpha}^\dagger_{\bf k} - v_{\bf k}\,\hat{\alpha}_{-{\bf k}}
\,,
\eea
with $u_{\bf k}^2=1+v_{\bf k}^2= [1+(\epsilon_{\bf k}+n\,U_0)/ \omega_{\bf k}]/2$,
$\epsilon_{\bf k}=k^2/2\,m$ and $\omega_{\bf k}^2= \epsilon_{\bf k}^2 +2 n U_0\,\epsilon_{\bf k}$. 
The stress tensor is given by Eq.\ (\ref{stress_uniform}) and the only contribution that gives nonzero matrix elements when acting on the ground state is that which creates two excitations, which is given by 
\be
(\hat{T^0}_\qq)_{xy} =- \sum_\pp \frac{p_x p_y}m \,     \left( u_{\pp-\qq}\,v_\pp + u_\pp\,v_{\pp-\qq}\right)    
{\hat
  \alpha}^\dagger_{\pp-\qq} {\hat \alpha}^\dagger_{-\pp} \,,
\ee
where we have also made use of the fact that the gravitational wave is transverse, and therefore $q_x=q_y=0$.
 
 Inserting the expressions above for the matrix elements of the stress tensor operator and the excitation energies into the general result (\ref{paraFT}), one finds 
 at zero temperature, and when analytically continued to a real frequency $\omega$, the result
\begin{widetext}
 \bea
\chi_{xy,xy}(\qq, \omega)
 &=&
   \int \, \frac{d\mathbf{p}}{(2\pi)^3}
   \,\frac{p_x^2\,p_y^2}{m^2}
   \,\left( u_{\pp+\qq}\,v_\pp + v_{\pp+\qq}\,u_\pp\right)^2
%\nonumber\\
%&~&
%\times
\,
     \Big[
       \frac{1}{\omega+\rmi\eta-(\omega_{\mathbf p+\mathbf q}+\omega_{\mathbf p})} 
        -
        \frac{1}{\omega+\rmi\eta+(\omega_{\mathbf p+\mathbf 
        q}+\omega_{\mathbf p})}
      \Big]
\,.
\nonumber\\
\label{bec_responseT=0}
\eea
\end{widetext}
Equation
(\ref{bec_responseT=0}) shows that the response does not vanish even for $q=0$.   It is not possible for a gravitational wave, which is transverse, to excite a single Bogoliubov excitation because the latter is longitudinal.  It is relevant to
stress that, while the contribution from states with two excitations to the transverse current-current response
  function vanishes in the long wavelength limit, the corresponding contribution to the
  stress-tensor--stress-tensor response remains nonzero, due to a
different sign in the Bogoliubov factors inside the parenthesis in
(\ref{bec_responseT=0}).   This is a specific example of the general result given in \mbox{Sec.\ \ref{sec:coupling}} and is a consequence of the fact that the stress tensor does not obey a conservation law when there are interactions.

The integral in Eq.\ (\ref{bec_responseT=0}) is ultraviolet divergent.
This is due to the fact that we have used an effective low energy theory to calculate a quantity that cannot be expressed in terms of the constants in the theory.  However, the response at frequencies with a magnitude of order $nU_0$ or less may found by subtracting from the response function its zero  frequency value.  For simplicity we consider the long-wavelength limit, $q\rightarrow 0$ and find
 \bea
\chi_{xy,xy}(\qq, \omega)&-&\chi_{xy,xy}(\qq, 0)
 =
  \omega^2 \int \, \frac{d\mathbf{p}}{(2\pi)^3}
   \,\frac{p_x^2\,p_y^2}{m^2}
\nonumber\\
&\times&
   \,\frac{\left(2 u_{\pp}\,v_\pp \right)^2}{2\omega_\pp} \Big[
       \frac{2\omega_\pp}{(\omega+\rmi\eta)^2-4\omega_{\mathbf p}^2}+\frac1{2\omega_\pp}
            \Big]\,.
\nonumber\\
\label{bec_responseT=0Subtracted}
\eea
The integral may be performed analytically, but the result gives little insight.  The imaginary part is simple, and is given by
\be
{\rm Im}\chi(\qq, \omega)=-\frac{1}{15\pi^2}\frac{(nU_0)^2}{\omega}\frac{p(\omega)^5}{2nmU_0+p^2(\omega)},
\ee
 where
 \be
 p(\omega)=\left(\frac{\sqrt{(4mnU_0)^2+m^2\omega^2}-4mnU_0}2\right)^{1/2}.
 \ee
At frequencies much less than $nU_0$ this varies as $\omega^4$ while for frequencies much larger than $nU_0$ (but still small enough for the low-energy theory to be valid) it varies as $\omega^{1/2}$.

\subsection{Fermionic superfluid}

In this section we compute the response, at zero temperature, of a
superfluid made up of pairs of fermions in two internal states, which we refer to as up and down. The contribution to the stress tensor operator from
the kinetic energy may be written as
\bea
\hat{T}_{ij}(\mathbf{q})
 &=& \frac{1}{m}
     \,\sum_p ({p}+\frac{{q}}{2})_i
               \,({p}+\frac{{q}}{2})_j
\nonumber\\
&~&~~~~~\times
             (\hat{a}^\dagger_{\mathbf{p}+\mathbf{q}\,\uparrow}
                    \,\hat{a}_{\mathbf{p}\,\uparrow}
               + \hat{a}^\dagger_{-(\mathbf{p}+\mathbf{q})\,\downarrow}
                    \,\hat{a}_{-\mathbf{p}\,\downarrow})
\,.
\eea
There will generally be in addition a contribution from the interaction energy but we ignore this since, for weak coupling, it is small compared with that from the kinetic energy.
We shall assume the superfluid to be of the BCS type,  with s-wave pairing between two spin states, denoted by ``up'' and ``down''. The elementary excitations are linear combinations of particles and holes, which are destroyed by operators $\hat\alpha_\mathbf{k}$, $\hat\beta_\mathbf{k}$ and created by the Hermitian conjugate operators.  The particle creation and annihilation operators are related to these by the canonical transformation
\bea \label{bog}
a_{\mathbf{k}\,\uparrow} &=& u_\mathbf{k}\,\hat\alpha_\mathbf{k} 
                   + v_\mathbf{k}\,\hat\beta^\dagger_{-\mathbf{k}}\, ,
\nonumber \\
a_{-\mathbf{k}\,\downarrow} &=& u_\mathbf{k}\,\hat\beta_{-\mathbf{k}} 
                     - v_\mathbf{k}\,\hat\alpha^\dagger_{\mathbf{k}}
\,,
\end{eqnarray}
where $u_q^2=1-v_q^2=1/2\,(1+\xi_q/E_q)$, $\xi_q=\epsilon_q-\mu$, and
$E_q= \sqrt{\Delta^2 + \xi_q^2}$ is the energy of an excitation, with $\Delta$
being the energy gap.  For simplicity, we restrict ourselves to zero temperature, and therefore the contribution to the stress tensor operator coming from the kinetic energy is 
\bea
 \hat{T}_{ij}(\mathbf{q})
 &=& \frac{1}{m} \, \sum_p ({p}+\frac{q}{2})_i\,({p}+\frac{q}{2})_j
            \, (u_{p+q}\,v_p + v_{p+q}\,u_p)
\nonumber\\
&~&~~~~~\times
            \, (\hat{\alpha}^\dagger_{\mathbf{p}+\mathbf{q}}
                                      \,\hat{\beta}^\dagger_{-\mathbf{p}}
               + \hat{\beta}_{-(\mathbf{p}+\mathbf{q})}
                                \,\hat{\alpha}_{\mathbf{p}}).
\,,
\eea
where we have neglected terms like $\hat{\alpha}^\dagger\,\hat{\beta}$ and $\hat{\alpha}\,\hat{\beta}^\dagger$ that vanish when acting on states with no excitations present.  For weak coupling, the contributions to the stress tensor operator from the interaction energy will be small compared with those from the kinetic energy, and we shall neglect them.

As before, the diamagnetic term is proportional to the pressure of the
fluid. In this section we consider the first term of the dispersion to
be the same as the one for the free fermion case, by neglecting the
effect of the interaction on the pressure. Therefore, the dispersion
relation reads 
\bea
\omega^2
   &\approx& 
      c^2\,q^2 +  \frac{32\pi}{5}Gmn\frac{ v_F^2 }{c^2} 
              + \frac{16\pi G}{c^2}\,\chi_{xy,xy}
\,. 
 \eea
The response function can be found by inserting the expressions for
the matrix elements into the general expression (\ref{paraFT})
\bea \label{T_BCS}
\chi_{xy,xy}(\mathbf{q},\omega)
&=& 
      \,\sum_{\mathbf{p}} 
      \frac{{p}_i\,{p}_j\,{p}_k\,{p}_l}{m^2}
      \, (u_{p+q}\,v_p + v_{p+q}\,u_p)^2
\nonumber\\
&\times&
\,
      \big[
        \frac{1}{\omega-(E_{p+q}+E_p)+\rmi\,\eta} 
\nonumber\\
&~&~~~~
        -
        \frac{1}{\omega+(E_{p+q}+E_p)+\rmi\,\eta}
      \big]
\nonumber \\
&\approx& 
    \frac{4}{m^2}
    \sum_{\mathbf{p}}
       p_i\,p_j\,p_k\,p_l\,
      \frac{\Delta^2}{E_p\,( (\omega+\rmi\,\eta)^2 - 4\,E_p^2)}
\,,
\nonumber\\
\eea
where the final expression is valid for $q \ll \Delta/v_F$.  The imaginary
term is nonzero only for $\omega > 2\,\Delta$, since the minimum
energy of a single excitation is $\Delta$.

\begin{figure}[t!]
\centering \includegraphics[height=2.8in,clip]{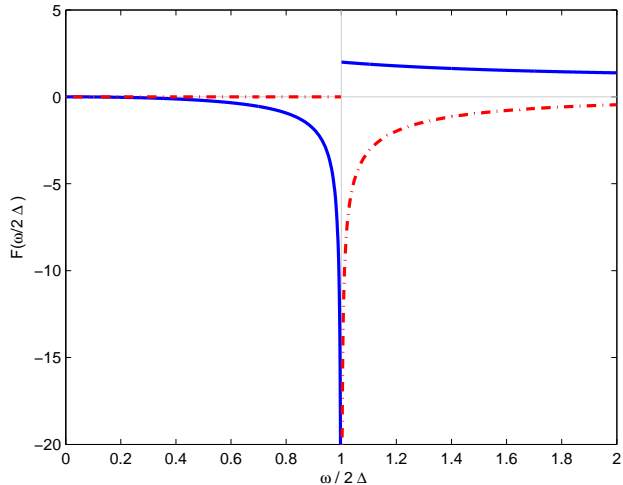}
\caption{Behavior of the function $F(\omega,\Delta)$,
  Eqs.~(\ref{Feq_minus}) and~(\ref{chi_BCS}). The full line is the real part of $F$ and the
  dashed line is the imaginary part.}
\label{F}
\end{figure}

The sum in (\ref{T_BCS}) gives a large contribution for $\omega \sim
2\,\Delta$, and far from this resonant frequency the response hardly
makes any contribution, since the self energy is proportional to
$G/c^4$.  More quantitatively, this sum can be transformed to an
integral, and this integral can be solved in spherical
coordinates. Integrating over the angles we have 
\bea 
\chi_{xy,xy}(\mathbf{q},\omega)
&=&
\frac{4\,\pi}{15}\, \frac{4}{(2\,\pi)^3\,m^2} \int_0^\infty d p \, p^6
\nonumber\\
&~&~~~~\times
\frac{\Delta^2}{E_p\,( (\omega+i\,\eta)^2 - 4\,E_p^2)} 
\,.
\nonumber\\
\eea
This integral does not converge, the problem being that an effective
low-energy theory has been used to calculate a quantity that has
important contributions from high-energy states.  To investigate the
low-frequency structure of the response function, we subtract from the
response function, its value for $\omega=0$. The integral converges
and consequently one may evaluate it putting $\Delta$ equal to its
value at the Fermi surface.
\begin{widetext}
\bea
\chi_{xy,xy}({\bf q}, \omega)-\chi_{xy,xy}({\bf q}, \omega=0)
&=&
 \label{T_BCS_partial}
~ \frac{1}{15}\,
  \frac{1}{2\,\pi^2\,m^2}
        \int_0^\infty d p \, p^6 
        \, \frac{\Delta^2\,\omega^2}
                {E_p^3\,( (\omega+\rmi\,\eta)^2 - 4\,E_p^2)}
\nonumber\\
&=& \frac{1}{5} \, n \,p_F\, v_F \, \Delta^2 \, \omega^2
     \int_{-\infty}^\infty d\xi \frac{1}{(\xi^2+\Delta^2)^{3/2}
                                \,( (\omega+\rmi\,\eta)^2-4\,(\xi^2+\Delta^2))}
\nonumber\\
&=& \frac{2}{5} \, n \,p_F\, v_F\,
     F\left(\omega/2\,\Delta\right)
\,,
\eea
\end{widetext}
where, for $\omega < 2\,\Delta$
\bea\label{Feq_minus}
F(\Omega) &=& 1
                - \left(
                       \Omega\,\sqrt{1-\Omega^2}
                     \right)^{-1}
                      \,\cos^{-1}
                                \left( \sqrt{1-\Omega^2}\right)
\nonumber\\
            &=& 1- 2\,\frac{\phi}{\sin 2\,\phi}
\,,
\eea
with $\Omega=\sin\phi$, and for $\omega >2\,\Delta$
\bea \label{chi_BCS}
F(\Omega) &=& 1 + \left(
                       \Omega\,\sqrt{\Omega^2-1}
                     \right)^{-1}
\nonumber\\
 &~&~~~~\times
                     \left(
                      \,\sinh^{-1}
                                \left( \sqrt{\Omega^2-1}\right)
                         - \rmi\,\frac{\pi}{2} 
                     \right)
\,.
\eea
The real and imaginary part of $F$ are plotted in Fig.~\ref{F}.

This calculations represents the simplest approximation for the
response, but they do not take into account residual interactions
between excitations.  Such interactions are important for the
collective behavior and lead, e.g., to the Bogoliubov--Anderson sound
mode \cite{Bogoliubov,Anderson}, which represents a density wave in the condensate.  However, the
effects of the residual interaction on the stress-tensor response function are
expected to be less dramatic since the it is
transverse, not longitudinal.  The present case is more analogous to
excitons in superconductors \cite{martin} and pairing vibrations in
atomic nuclei \cite{bohrmottelson}, where the momentum dependence of
the interaction plays a crucial role.

\section{Concluding remarks}

\label{sec:conclusion}

In this paper, we have developed a general framework for studying the
interaction of a weak gravitational wave with matter.  A virtue of the
approach is that it separately clearly the effects of general
relativity from the problem of solving the many-body problem for the
matter. The matter gives a self energy to the propagator of the
gravitational wave. This self energy has contributions analogous to
the paramagnetic and diamagnetic contributions to the self energy of
an electromagnetic wave in matter.  The contribution corresponding to
the paramagnetic term is proportional to the
stress-tensor--stress-tensor correlation function for the
matter. Because the stress tensor operator is not a conserved
quantity, except for non-interacting particles, this correlation
function does not in general vanish in the long wavelength limit for
nonzero frequency, and we illustrated this by explicit calculations
for a Bose--Einstein condensate and a BCS superfluid.  The general
formalism in this paper makes for a very simple derivation of the
dispersion relation for gravitational waves in astrophysical plasmas.

There are a number of possible directions for future work.  In this paper we have considered only an infinite medium, and one could extend the treatment to take into account the effect of boundaries.  Another application is to systems, like metals and superconductors, in which the Coulomb interaction plays a key role.  The formalism may also be used to establish  the relationship
between, on the one hand, the microscopic theory in terms of particles
and their interactions and, on the other hand, elastic theory which
has been commonly used to discuss the response of gravitational wave
antennas.

\begin{acknowledgments}
During the course of this work we have benefitted from many fruitful
discussions with colleagues, including James Bardeen, Michael Bradley,
Gert Brodin, Hael Collins, Poul Henrik Damgaard, Benny Lautrup,
Mattias Marklund, Savvas Nesseris, Niels Obers, Poul Olesen, Kip
Thorne, and Zhenhua Yu. The visits of AC to the Niels Bohr
International Academy were supported in part by the Intercan Network,
the Rosenfeld Foundation, and NORDITA.
\end{acknowledgments}


\begin{thebibliography}{10}
\bibitem{foot1} We adopt the convention that Greek indices run over the values 0,1,2  and 3, while Roman indices run over the spatial indices 1,2 and 3.
\bibitem{foot2} We shall use the word ``matter'' as a shorthand for all degrees of freedom other than those of the gravitational field.
\bibitem{foot3} We employ units in which $\hbar$ and the Boltzmann constant are unity.
\end{thebibliography}
\end{document}